\documentclass[sigconf,dvipsnames,screen]{acmart}

\AtBeginDocument{%
  \providecommand\BibTeX{{%
    \normalfont B\kern-0.5em{\scshape i\kern-0.25em b}\kern-0.8em\TeX}}}

\setcopyright{acmcopyright}
\acmPrice{} 
\acmDOI{10.1145/3540250.3549131}
\acmYear{2022}
\copyrightyear{2022}
\acmSubmissionID{fse22main-p539-p}
\acmISBN{978-1-4503-9413-0/22/11}
\acmConference[ESEC/FSE '22]{Proceedings of the 30th ACM Joint European Software Engineering Conference and Symposium on the Foundations of Software Engineering}{November 14--18, 2022}{Singapore, Singapore}
\acmBooktitle{Proceedings of the 30th ACM Joint European Software Engineering Conference and Symposium on the Foundations of Software Engineering (ESEC/FSE '22), November 14--18, 2022, Singapore, Singapore}

\usepackage[utf8]{inputenc}
\usepackage{url}
\usepackage{colortbl}
\usepackage{caption}
\usepackage{balance}
\usepackage{xspace}
\usepackage[caption=false]{subfig}
\usepackage{graphicx}
\usepackage{verbatim}
\usepackage{bold-extra}
\usepackage{amsmath}
\usepackage{multirow}
\usepackage{listings}
\usepackage{boxedminipage}
\usepackage{soul}
\usepackage{enumitem}
\usepackage[normalem]{ulem}
\setlist{nolistsep,leftmargin=.5cm}
\usepackage{booktabs}
\usepackage{tabularx}
\usepackage{svg}
\usepackage{calc}
\usepackage{float}
\usepackage{multirow}
\usepackage[normalem]{ulem}
\useunder{\uline}{\ul}{}
\usepackage{amsmath}
\usepackage[most]{tcolorbox}
\usepackage{caption}
\usepackage{microtype} 

{

\newcommand{\ie}{\textit{i.e.},\xspace}
\newcommand{\eg}{\textit{e.g.},\xspace}
\newcommand{\etc}{\textit{etc.}\xspace}
\newcommand{\etal}{\textit{et al.}\xspace}

\newcommand{\tool}{{\sc{Burt}}\xspace}
\newcommand{\app}{\tool}
\newcommand{\ap}{\tool}

\newcommand{\itr}{{\sc{Itrac}}\xspace}
\newcommand{\itrac}{\itr}

\newcommand*{\img}[1]{%
    \raisebox{-.2\baselineskip}{%
        \includegraphics[
        height=\baselineskip,
        width=\baselineskip,
        keepaspectratio,
        ]{#1}%
    }%
}

\usepackage{tikz}

\definecolor{bug_red}{rgb}{.84,.23,.29}

\definecolor{info-needed-color}{rgb}{1,.8,.12}

\newcommand{\emphquote}[1]{{\emph{`#1'}}\xspace}

\newcommand{\crashscope}{{\scshape{CrashScope}}\xspace}
\newcommand{\euler}{{\scshape{Euler}}\xspace}
\newcommand{\fusion}{{\scshape{Fusion}}\xspace}
\newcommand{\ebug}{{\scshape{EBug}}\xspace}

\newcommand{\subj}{\texttt{\small subject}}
\newcommand{\act}{\texttt{\small action}}
\newcommand{\obj}{\texttt{\small object}}
\newcommand{\prep}{\texttt{\small preposition}}
\newcommand{\objtwo}{\texttt{\small object2}}

\begin{document}

\title{
	Toward Interactive Bug Reporting for (Android App) End-Users
}

\author{Yang Song}
\affiliation{%
	\institution{College of William \& Mary}
	\country{USA}
}

\author{Junayed Mahmud}
\affiliation{%
	\institution{George Mason University}
	\country{USA}
}

\author{Ying Zhou}
\affiliation{%
	\institution{University of Texas at Dallas}
	\country{USA}
}

\author{Oscar Chaparro}
\affiliation{%
	\institution{College of William \& Mary}
	\country{USA}
}

\author{Kevin Moran}
\affiliation{%
	\institution{George Mason University}
	\country{USA}
}

\author{Andrian Marcus}
\affiliation{%
	\institution{University of Texas at Dallas}
	\country{USA}
}

\author{Denys Poshyvanyk}
\affiliation{%
	\institution{College of William \& Mary}
	\country{USA}
}

\renewcommand{\shortauthors}{Song et al.}

\begin{abstract}

Many software bugs are reported manually, particularly bugs that manifest themselves visually in the user interface. 
End-users typically report these bugs via app reviewing websites, issue trackers, or in-app built-in bug reporting tools, if available. 
While these systems have various features that facilitate bug reporting  (\eg textual templates or forms), they often provide \textit{limited} guidance, concrete feedback, or quality verification to end-users, who are often inexperienced at reporting bugs and submit low-quality bug reports that lead to excessive developer effort in bug report management tasks.

We propose an interactive bug reporting system for end-users (\ap), implemented as a task-oriented chatbot. 
Unlike existing bug reporting systems, \ap provides guided reporting of essential bug report elements (\ie the observed behavior, expected behavior, and steps to reproduce the bug), instant quality verification, and graphical suggestions for these elements. 
We implemented a version of \ap for Android and conducted an empirical evaluation study with end-users, who reported 12 bugs from six Android apps studied in prior work. 
The reporters found that \ap's guidance and automated suggestions/clarifications are useful and \ap is easy to use. 
We found that \ap reports contain higher-quality information than reports collected via a template-based bug reporting system. 
Improvements to \ap, informed by the reporters, include  support for various wordings to describe bug report elements and improved quality verification. Our work marks an important paradigm shift from static to interactive bug reporting for end-users. 

\end{abstract}



\maketitle


\vspace{0.2cm}
\section{Introduction}

Bug report management is an important and costly software engineering activity. While certain types of bugs can be reported automatically via a known oracle (\eg crashes), recent studies have illustrated that more than half of the bugs reported in open source software relate to functional problems with no automatically identifiable oracle~\cite{Tan2014} and, hence, must be reported manually.
High-quality bug reports are essential for bug triage  and resolution 
and they are expected to describe \textit{at minimum} the observed (incorrect) behavior (\textbf{OB}), the steps to reproduce the bug (\textbf{S2Rs}), and the expected (correct) software behavior~(\textbf{EB})~\cite{Bettenburg2008a, Laukkanen2011, Zimmermann2010}.


One of the main difficulties that contributes to quality issues in end-user bug reporting is the \textit{knowledge gap} between end-users and developers~\cite{Moran2015, Huo2014}. That is, there is often a gap between what end-users \textit{know} and what developers \textit{need}~\cite{Zimmermann2010}, generally due to the fact that users are both unfamiliar with the internals of the software and with the explicit types of information that are important for developers (\eg the OB, EB, and S2Rs). 

Most current reporting systems are not designed to address the above-mentioned knowledge gap between end-users and developers.
In particular, current systems are typically lacking along two important dimensions: (1) they offer \textit{limited guidance} related to \textit{what} needs to be reported and \textit{how} it needs to be reported; and (2) 	no \textit{feedback} is offered to reporters on whether the information they provided is correct or complete. In consequence, given the \textit{static nature} of these bug reporting interfaces, the burden of providing high-quality information rests on the reporters.
 

We posit that an \textit{interactive} reporting solution can help to bridge the developer--end-user knowledge gap. 
Inspired by prior work on question/answering systems for debugging~\cite{Ko2008}, we argue that a conversational agent (\ie a chatbot) can successfully guide end-users through the reporting process, while offering interactive suggestions and instant quality verification. 

In this paper, we introduce and evaluate a task-oriented dialogue system for \textbf{BU}g Repo\textbf{RT}ing (or \tool) that is capable of providing instant feedback for each element of a bug description (\ie OB, EB, and S2Rs), while actively guiding corrections, where needed. 
\tool combines novel and state-of-the-art techniques for dynamic software analysis, natural language processing, and automated report quality assessment. 
We designed and developed the current version of \ap to work for Android apps, but its architecture is platform-agnostic and it can be instantiated, with some engineering effort, for other types of GUI-based applications (\eg web-based, desktop, or iOS-based).
In particular, \tool constructs a graph of program states using both crowdsourced app usage data and automated GUI-based exploration techniques. 
The chatbot then parses and interprets end-user descriptions of various bug report elements by matching them to states and transitions in the constructed graph, and produces graphical suggestions regarding information that is likely to be reported (\eg the next S2Rs). 
Additionally, \tool recognizes when end-users provide incomplete or ambiguous information and suggests improvements or clarifications to the users. 
Traditional task-oriented chatbots typically have direct access to a structured and  easily parseable knowledge-base~\cite{chatbot-arch}.
In contrast, \ap is more complex, as it reconciles high-level descriptions provided by end users and matches these to technical program information, bridging the end-user to developer knowledge gap. 

We evaluated \tool empirically, asking 18 end-users, with various levels of prior bug reporting experience, to report 12 bugs from six Android apps using a prototype implementation of \tool.
We found that the guidance and automated suggestions/clarifications made by the chatbot were accurate, useful, and easy to use, and the collected bug reports are high-quality. 
We asked 18 additional end-users to report the same bugs with a template-based bug reporting system (\itr) and compared the quality of these reports to those reported with \ap. 
\ap reports have fewer incorrect and missing S2Rs than the \itr reports. 
We also found that \ap helps novice bug reporters provide more correct steps, and experienced reporters avoid missing steps. 


In summary, the contributions of this paper are as follows: 
\begin{itemize}
	\item{\tool, the first task-oriented, conversational agent that supports end-users in reporting bugs (currently for Android apps), with features such as automated suggestions, real-time feedback, prompts for information clarification, and graphical cues.} 
	\item{The results of an empirical evaluation involving 36 end-users that investigates user experiences, preferences, and attributes of interactive bug reporting with \tool, as well as the quality of the resulting bug reports. }
	\item{A replication package~\cite{repl_pack} that contains a complete implementation of \tool, \tool's app usage/execution data, code and data about \ap's evaluation, and documentation that enables the verification and validation of our work and future research in the topic of bug reporting systems. The package also includes a video illustrating \ap.
	}
\end{itemize}

Our work opens the door to a new way of thinking about end-user bug reporting, using conversational agents, shifting the state of the art from \textit{static} to \textit{interactive} bug reporting. 
While \tool is a prototype, we expect that it will serve as the foundation for a new class of interactive bug reporting systems, combining elements of existing static systems with features of conversational agents~\cite{conversational-ai}.

\section{B{\large URT}: A Chatbot for Bug Reporting}

We propose a task-oriented chatbot for \textbf{BU}g Repo\textbf{RT}ing (\tool). \ap offers a variety of features for interactive bug reporting such as the ability to (i) guide the user in reporting essential bug report elements, (ii) check the quality of these elements, (iii) offer instant feedback about issues, and (iv) provide graphical suggestions.

\ap is designed to collect three key elements for developers during bug triage and resolution~\cite{Zimmermann2010,Laukkanen2011,Sasso2016}: the \textit{observed behavior} (\textbf{OB}), the \textit{expected  behavior} (\textbf{EB}), and the \textit{steps to reproduce} the bug (\textbf{S2Rs}).
\ap collects these from the user through a dialogue and generates a web-based bug report containing textual descriptions for these elements with attached screen captures of the system.

\ap's design consists of three main components, 
inspired by the typical architecture of task-oriented dialogue systems~\cite{conversational-ai}, which adapt techniques from automated program analysis and natural language processing to facilitate bug reporting. 
\ap's \textbf{Natural Language Parser (NL)} parses the relevant information from end-user responses to the chatbot.
The \textbf{Dialogue Manager (DM)} dictates the structured conversation flow for \ap's reporting process and handles the presentation of multi-modal (\eg screenshots and text) information to the user. 
Finally, the \textbf{Report Processing Engine (RP)} maps information parsed from user responses to various states in a program execution model for a given app in order to assess bug element quality. 
The current version of \ap is designed for Android apps and builds its execution model using a combination of automated app exploration and crowdsourced user traces. 
In this section, we present \ap 's components in detail. 

\begin{figure}[t]
	\centering{
		\includegraphics[width=\linewidth]{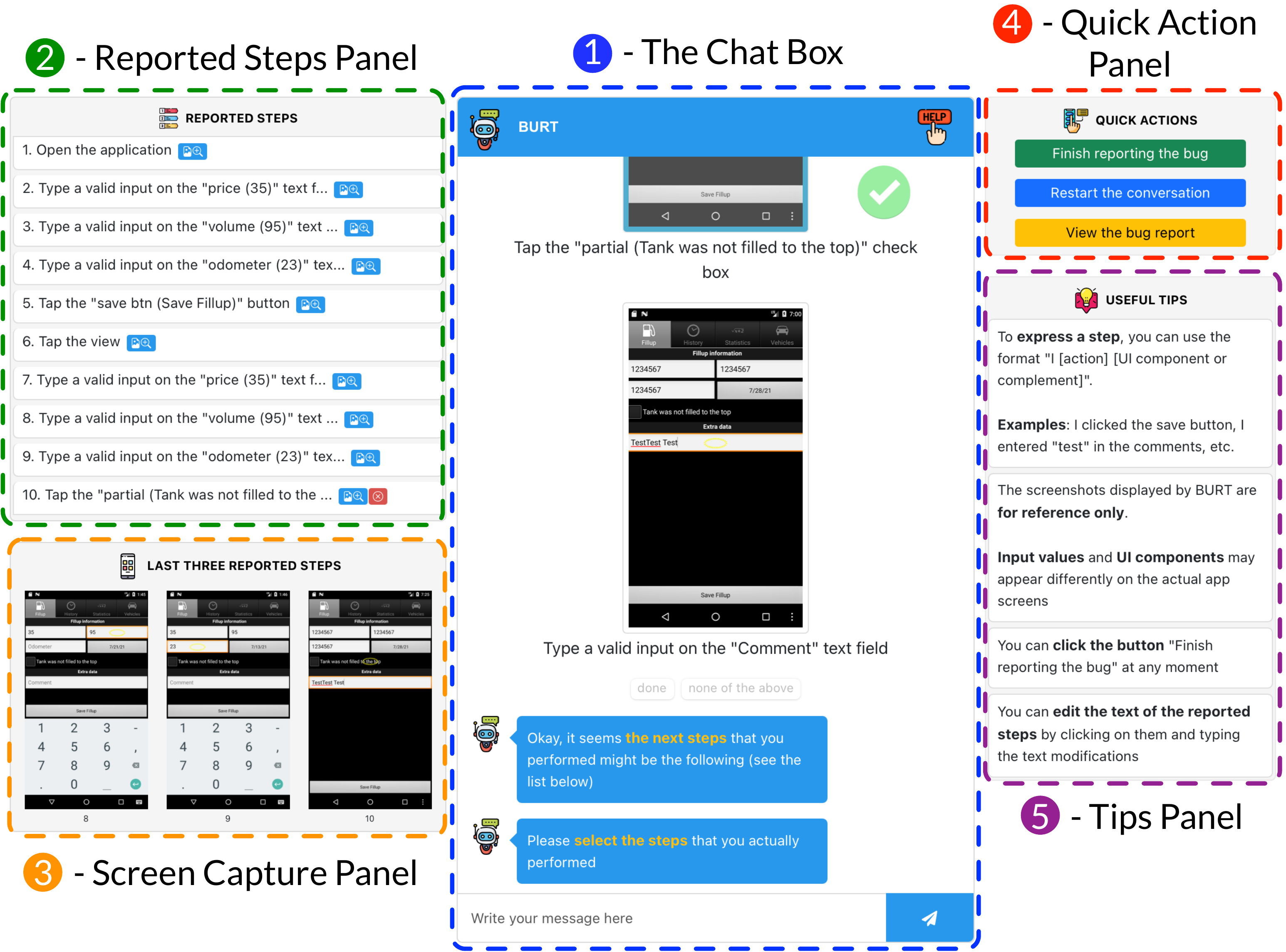}
		\caption{B{\footnotesize URT}'s graphical user interface}
		\label{fig:burtgui}
		\vspace{-0.5cm}
	}
\end{figure}

\subsection{Graphical User Interface (GUI)}

We designed \ap as a web-based application that includes both a standard chatbot interface along with additional visual components as illustrated in Fig. \ref{fig:burtgui}. The \textit{Chat Box} \img{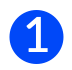} allows the end-user to provide textual descriptions of the OB, EB, and S2Rs as well as interact with the graphical information that \ap displays (\eg recommendations of the next S2Rs via screenshots). The \textit{Reported Steps Panel} \img{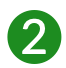} enumerates and displays the S2Rs that the user has reported. 
The textual description of the reported steps can be edited and the last reported step can be deleted, if the user makes a mistake and wishes to correct it. The \textit{Screen Capture Panel} \img{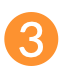} displays screen captures of the last three S2Rs. The \textit{Quick Action Panel} \img{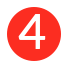} provides buttons to finish reporting the bug, restart the bug reporting session, and (pre)view the bug report being created -- these can be activated anytime. 
The \textit{Tips Panel} \img{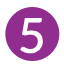} displays recommendations to end-users on how to use \ap and how to better express the OB, EB, and S2Rs. The tips change depending on the current stage of the conversation. 


\subsection{Natural Language Parser (NL)}

\ap parses the OB, EB, and S2R descriptions provided by end-users using dependency parsing via the Stanford CoreNLP toolkit~\cite{Manning2014}. This process obtains the tree of grammatical dependencies~\cite{StanfordDependencies} between words in a sentence and extracts the relevant words from the tree. This parsing technique is needed
by the Report Processing Engine to assess the quality of parsed bug report elements and to help direct the flow of conversation (see Sec.~\ref{sec:approach:matching}).

\ap first utilizes the heuristic-based approach introduced by Chaparro \etal~\cite{Chaparro:FSE17} to identify the type of a sentence (\eg conditional, imperative, or passive voice) for each message received from the user. This approach implements heuristics (based on dependency parsing and part-of-speech tagging~\cite{Manning2014}) to identify  discourse patterns in OB, EB, and S2R descriptions~\cite{Chaparro:FSE17}. Once the sentence type is identified, \tool executes a series of algorithms to extract the relevant words from the sentence, based on prior work on quality assessment of S2Rs~\cite{Chaparro:FSE19}. In essence, we implemented 16 parsing algorithms that traverse the grammatical trees~\cite{StanfordDependencies} of end-user sentences which have a different structure depending on the sentence type (\eg conditional or imperative). Each algorithm parses sentences of one type. All the 16 algorithms implemented for the different types of OB/EB/S2R sentences can be found in our online replication package~\cite{repl_pack}.

\ap parses a single sentence using the following format:

\noindent\hspace{1mm} \texttt{[subject] [action] [object] [preposition] [object2]}  
\noindent where the \subj ~is the actor (\eg the user or an app component) performing the \act, which is an operation or task (\eg tap, create, crash); the \obj ~is an ``entity'' directly affected by the \act, and \objtwo ~is another ``entity'' related to the object by the \prep.
 An ``entity'' is a noun phrase that may represent numeric/textual app input, domain concepts, GUI components, \etc 
Depending on the sentence, its type, and whether it describes an OB, EB, or S2R, the words (\eg the \subj, \prep ~or \objtwo) extracted from the entity are required or optional.

For example, for the Mileage Android app~\cite{mileage}, the OB sentence \textit{``The average fuel economy shows a NaN value''}, written in present tense, is parsed as \texttt{\small [average fuel economy] [shows] [NaN value]}. The EB sentence \textit{``fuel economy statistics should be calculated correctly''}, which uses the modal ``should'', is parsed as \texttt{\small [average fuel economy] [is] [calculated]}. The S2R sentence \textit{``Save the car fillup''}, written imperatively, is parsed as \texttt{\small [user] [saves] [car fillup]}. 

Some sentences describe a combination of OB, EB and S2Rs in a single phrase. For example, the sentence \textit{``The app stopped when I added a new time range''} describes both an OB and a S2R. 
This sentence is parsed by \tool as \texttt{\small [app] [stopped]} as the OB, and \texttt{\small[add] [new time range]} as the S2R. 
In this example, \ap extracts the S2R from the sentence as follows. First, it locates the adverb \textit{``when''} in the parsed grammatical tree, then it follows the relationship that leads to the verb \textit{``add''} for which \textit{``when''} is the adverbial modifier. Next, \ap locates the verb's nominal subject ``I'' and its direct object \textit{``time range''}. If these relationships do not exist in the tree, the sentence is not conditional, as expected. Otherwise, \ap extracts the verb \textit{``add''} as the \act~and the noun phrase \textit{``time range''} as the \obj. In the end, this sentence is parsed as the S2R: \texttt{\small[add] [new time range]}.

When  multiple sentences compose a single user message, \ap only parses the initial sentence. When \ap is unable to parse a user message (\eg because it cannot identify the \subj), it asks the user to rephrase the sentence. \ap's \textit{Tips Panel} \img{figures/5.png} and user guide suggests patterns to the user to phrase the OB, EB, and S2Rs.

\begin{figure}[t]
\centering{
\includegraphics[width=0.75\linewidth]{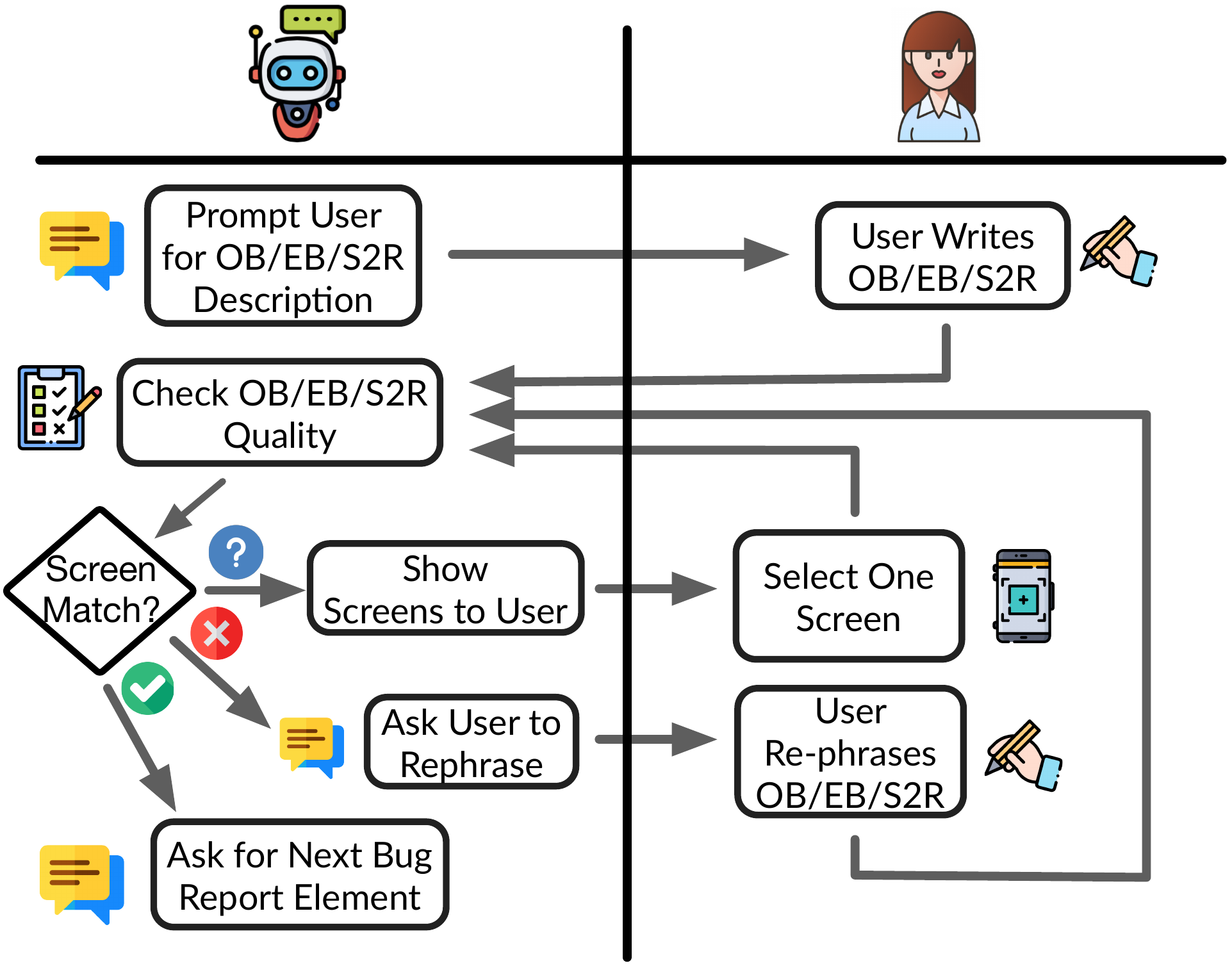}
\caption{\ap's dialogue flow for quality checking}
\label{fig:conversationflow}
\vspace{-0.3cm}
}
\end{figure}

\subsection{Dialogue Manager (DM)}
\label{subsec:dialouge-manager}

\ap's dialogue flow consists of three main phases: OB, EB, and S2R collection. \ap's dialogue is multi-modal in nature, and is capable of suggesting both natural language and graphical elements, such as screenshots, to help guide the user through the reporting process. 
The DM relies upon the RP engine to assess the quality of bug elements reported by end users (see Sec. \ref{sec:approach:matching}). 
While \ap's dialogue flow proceeds linearly to capture each bug element (the OB, EB, and S2Rs, in that order), the dialogue flow is similar for all elements. There are two main dialogue flows that \ap navigates: (i) performing quality checks on written bug report elements (applies to all bug elements), and (ii) automated suggestion of S2Rs (for S2Rs only). Next we describe these two main dialogue flows.

\subsubsection{Dialogue Flow for Bug Element Quality Checks (OB/EB/S2R)} 
\label{sec:approach:flow:quality}

Before the dialogue begins, a user must select the target app by clicking on its icon. Then, \ap's dialogue flow for quality checking, illustrated in a modified swimlane diagram in Figure~\ref{fig:conversationflow}, is initiated, starting with the OB. To begin the quality checking process, \ap prompts the user to provide the bug element (OB/EB/S2R). 
\ap automatically parses the description of the element and the RP engine verifies its quality (see Sec. \ref{sec:approach:matching}).

If the OB/EB/S2R is matched to an app screen from \ap's execution model (see Sec.~\ref{sec:approach:execution_model}), \ap asks the user for confirmation of the matched screen. 
If the user confirms, \ap proceeds to the next phase of the conversation (\eg asking for the EB or next S2Rs), otherwise, \ap asks the user to rephrase the bug element.

If there are no app screen matches, \ap informs the user about the issue and asks her to rephrase the OB/EB/S2R. 
Once the user provides a new description, the quality verification procedure is re-executed. 
If there are \textit{multiple} matches, \ap provides a list of up to five app screenshots (derived from the app execution model) that match the description. 
The user can then inspect the app screens and select the one that she believes best matches her description of the bug element. 
If none are selected, \ap suggests additional app screens if any. If the user selects one app screen, \ap saves the bug element description and screen, and proceeds to collecting the next bug element. After three unsuccessful attempts to provide a high quality OB description, \ap records the (last) provided OB description for bug report generation. This process proceeds for each bug element starting with the OB. S2Rs are treated slightly differently since \ap can also \textit{predict} S2Rs as we describe next.


\begin{figure}[t]
		\vspace{-0.2cm}
		\centering{
		\includegraphics[width=0.70\linewidth]{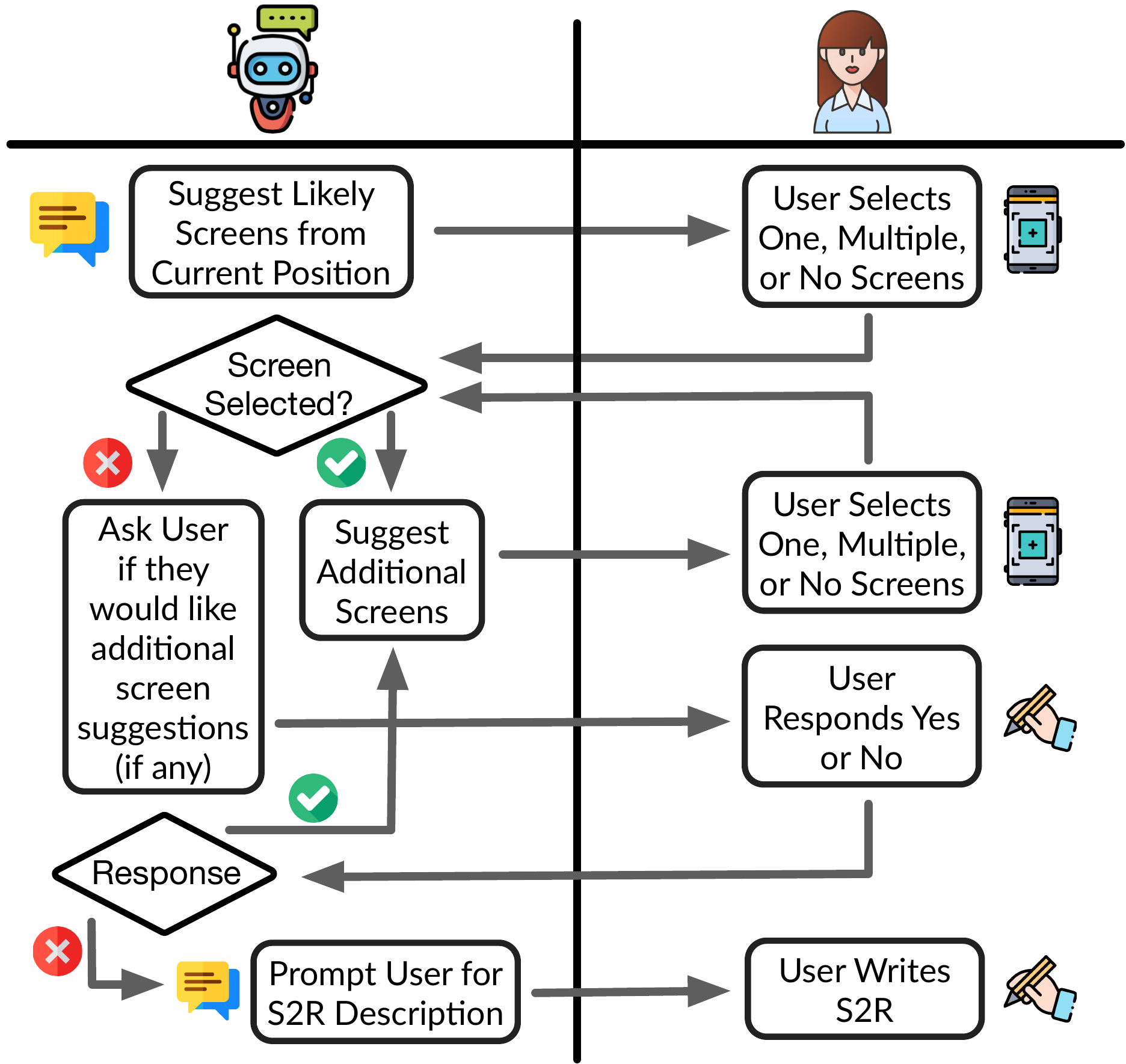}
		\caption{Dialogue Flow for S2R Predictions}
		\label{fig:s2r_conversationflow}
	}
\end{figure}

\subsubsection{Dialogue Flow for Suggesting S2Rs} 
\label{sec:approach:flow:quality}

B{\footnotesize URT} suggests potential next S2Rs that the user may have performed during actual app usage, depending on the last reported step and the user-selected screen that is having the problem, \ie the OB screen. Figure~\ref{fig:s2r_conversationflow} illustrates this process. This dialogue flow uses a predictive algorithm that uses \ap's execution model (see Sec. \ref{sec:approach:next_steps}). The suggestions are displayed as a list of app screens, each screen representing a S2R. Each S2R in the list displays the screen capture with a textual description placed below the image. The screen capture is visually annotated with a yellow oval highlighting the GUI component (\eg a button) executed by the step. The user can select none, one, or multiple of the suggested S2Rs. When a S2R is selected, \ap suggests additional S2Rs if any. When none are selected and \ap has more suggestions, \ap asks the user if she wants to get more suggestions. If so, \ap displays them. Otherwise, \ap prompts the user to describe the next S2R.

\subsubsection{Collecting Input Values}
User input from type-like steps (\eg ``\textit{I entered 5 gallons}'') are extracted by \ap from the \obj~or \objtwo~of the parsed S2Rs, by identifying literal values or quoted text. If the input value is missing or
generic (\ie not a literal or ``text''), \ap prompts the user to provide the input. This is only activated if the matched S2R is confirmed by the user as a correct~S2R.

\subsection{Report Processing Engine (RP)}
\ap's RP Engine is composed of three sub-components: (i) the \textit{App Execution Model}, (ii) the \textit{Dialogue Quality Processor} which maps parsed bug report elements to app states from the model, 
and (iii) the \textit{S2R Response Predictor} which infers likely next S2Rs, given an existing set of S2Rs already mapped to the execution model. 

\vspace{0.2cm}
\subsubsection{App Execution Model} 
\label{sec:approach:execution_model}
The app execution model is a graph that stores sequential GUI-level app interactions (\eg taps, types, or swipes performed on screen GUI components) and the app response to those interactions (\ie app  screens). These interactions and app responses are produced using two strategies: (1) by executing an automated systematic app exploration adapted from \crashscope's \textit{GUI-ripping Engine}~\cite{Moran2016,Moran:2017}, and (2) by recording (crowdsourced) app usage information from app end-users or developers. 
Both the systematic app exploration and app usage data are collected \textit{before} \ap is deployed for use.

\textbf{App Execution Model Data Collection.} 
This is \ap's plat-form-specific part and would be constructed differently for non-Android applications.
\ap uses a version of \crashscope's \textit{GUI-ripping engine}~\cite{Moran2016,Moran:2017} to generate app execution data in the form of sequential interactions. 
\crashscope enables dynamic analysis of Android apps that utilizes a set of systematic exploration strategies (\eg top-down and bottom-up) and has been shown to exhibit comparable coverage to other automated mobile testing techniques~\cite{Moran2016}. For a detailed description of the engine, we refer the  readers to Moran~\etal's previous work~\cite{Moran2016,Moran:2017}.
As in prior work~\cite{Gomez2013,Bernal:ICSE20,Cooper:ICSE21,Bernal:TSE22,Havranek:ICSE21}, we instantiate data collection by recording low-level app event traces using the \texttt{\small getevent}, \texttt{\small sendevent}, \texttt{\small uiautomator} utilities included in the Android OS and SDK.

Collecting crowdsourced user app usage data serves two main purposes: (1) increase the coverage of app states and screens in \ap's execution model;
 and (2) augment the model with scenarios that are common during normal app usage.
Section \ref{sec:approach:implemenation_limitations} describes the procedure that we implemented to collect the crowdsourced data.
Crowdsourced data collection leads to the same types of app events as the automatic app exploration does.


\textbf{App Execution Model Structure.} The execution model is a directed weighted graph $G = (V, E)$, where $V$ is the set of unique \textit{app screens} with complete GUI hierarchies \cite{AndroidLayout}, and $E$ is a set of \textit{app 	interactions} performed on the screens' GUI components. In this model, two screens with the same number, type, size, and hierarchical structure of GUI components are considered a single vertex. $E$ is a set of  unique tuples of the form $(v_x, v_y, e, c)$, where $e$ is an application event (tap, type, swipe, \etc) performed on a GUI component $c$ from screen $v_x$, and $v_y$ is the resulting screen right after the interaction execution. Each edge stores additional information about the interaction, such as the textual data input (only for \textit{type} events) and the interaction execution order dictated by the app usage (manual or automatic). The graph's starting node has only one outgoing interaction, which corresponds to the application launch. A GUI component is uniquely represented by a type (\eg a button or a text field), an identifier, a label (\emphquote{OK} or \emphquote{Cancel}), and its size/position in the screen. Additional information about a component is stored in the graph, for example, the component description given by the developer, the parent/children components, and an annotated screen capture of the app highlighting the GUI component being interacted with. The screen captures are used in the screen suggestions made by \ap (see Sec. \ref{sec:approach:next_steps}).

The graph edges have a weight which indicates the likelihood of a given app interaction represented as a state transition. The weights are utilized by the \textit{S2R Response Predictor} (see Sec. \ref{sec:approach:next_steps}), which aims to suggest S2Rs that end-users would perform when normally using given app features. To enable accurate predictions, \ap assigns higher weights to interactions executed by humans than those executed automatically by \crashscope. To accomplish this, \ap sets the weight of an edge to the number of times it was executed in the collected usage data. If an edge is not executed by a human, but was executed by \crashscope's systematic exploration, then edge weight is set to one, even if \crashscope executes the same interaction multiple times. 
While this weight assignment scheme is straightforward, it proved to be effective (see Sec.~\ref{sec:results}). 

\subsubsection{Dialogue Quality Processor}
\label{sec:approach:matching}
Based on prior work~\cite{Chaparro:FSE19}, \ap's quality definition is based on the ability to match a textual bug description (OB, EB, or S2R) to the screens (states) and interactions (edges) of the execution model. 
A textual description is considered to be high-quality if it can be precisely matched to the execution model, otherwise it is deemed low-quality. 
This definition and \ap's dialogue features that prompt users to improve low-quality descriptions aim to reduce the knowledge gap between the reporters, who are unfamiliar with app internals and may not know how to express a bug, and developers, who define and implement the vocabulary captured in \ap's execution model.


\textbf{Assessing OB Quality.} \ap first builds a query to the app execution model by concatenating the non-empty elements from the parsed description, namely the \subj, \act, \obj, and \objtwo. Then, it preprocesses the query using lemmatization~\cite{Manning2014} and attempts to retrieve all matching GUI components via an adapted version of the matching procedure proposed by prior work
~\cite{Chaparro:FSE19}. This procedure computes the similarity score between the query and the elements from a GUI component, namely the component label, the description, and the ID specified by the original developer. The similarity is computed based on a normalized length of the longest common substring between query and the component elements. If such similarity is greater than or equal to 0.5, then there is a match, otherwise there is a mismatch. 
If the initial query does not match an app screen, \ap runs a different query by using only the \subj, since, based on our experience, it may indicate a key GUI component that is directly related with a bug.

\ap keeps a list of the app screens with at least one matching GUI component. Such a list is sorted in increasing order by the distance between the starting state in the execution model and the matched state. If this list is empty, it means the OB description does not use vocabulary from the app screens and needs to be rephrased. If this list contains only one element, it is used to show the user the potential buggy app screen, which the user has to confirm as correct or incorrect. Otherwise, if the list contains multiple elements, it is used to display the possible buggy app screens so that the user decides the appropriate screen. The selected OB screen by the user is tracked in the execution model and is used for (1) EB description matching, (2) the prediction of the next S2Rs, and (3) asking the user if the last provided S2R is the last step to replicate the bug.

\textbf{Assessing EB Quality.} \ap performs the matching approach described above using the parsed EB description against the OB screen confirmed by the user. \ap assumes the OB screen is the one that should work correctly, therefore, it attempts to match the EB description to it. If the user did not select an OB screen, the EB matching is bypassed
 and the EB description is saved for generating the bug report. 
If the EB description does not match the OB screen, it means the vocabulary used in the EB description is different from the OB screen, and the EB description should be rephrased. However, 
rather than prompting the user to rephrase it, \ap asks the user if the OB screen is the one that should work correctly. 

\textbf{Assessing S2R Quality.} \ap adapts the step resolution/matching algorithm proposed by Chaparro \etal~\cite{Chaparro:FSE19} and performs exploration of the execution model driven by the matching of the reported S2Rs.
By default, \ap assumes the first S2R performed by a user is launching the app and the current graph state is set to be the first app screen that results from this operation. 

For a provided S2R description, starting from the current state, \app traverses the graph in a depth-first manner and performs step resolution on each state.  Step resolution is the process of determining the most likely app interactions that the S2R refers to in a particular state (\ie app screen). The result is a set of \textit{resolved
interactions} for the S2R on the selected states. If the S2R resolution
fails for these states (either with a mismatch or a multiple-match result), then it means that either: (1) there
are app states 
not present in the execution model, or (2) the S2R description is of low-quality. 

The \textit{resolved interactions} are matched against the interactions (\ie the edges) from the graph, by matching their source state $v_x$, the event $e$, and the component $c$. 
If a pair of interactions match, then they are considered to be the same interaction. 
The matching returns a set of interactions from the graph that match the resolved ones. 
If this set is empty, it means that the resolved interactions were not covered by the app exploration and the quality assessment returns a low-quality result with a mismatch. 
If the reason for the mismatch is because of multiple-component or -event match (\ie the S2R description matches multiple GUI components or map to multiple events), \ap considers the S2R as ambiguous, and \ap indicates that the S2R's \act  ~corresponds to multiple events, or the \obj ~or \objtwo ~match multiple GUI components. 
If there is a no-match, \ap specifies the problematic vocabulary from the S2R elements: \act, \obj, \objtwo, or any combination of these.

Otherwise, if the set of \textit{resolved interactions} is not empty, \ap proceeds with selecting the most relevant interaction that corresponds to the S2R description, by selecting the one whose source state is the nearest to the current execution state in the graph.


\subsubsection{S2R Response Predictor}
\label{sec:approach:next_steps}

B{\footnotesize URT} predicts the next S2Rs that a user may have performed in practice. 
The prediction is  executed during the following dialogue scenarios (see Fig. \ref{fig:s2r_conversationflow}): (1) when an OB screen from the execution model has been selected/confirmed by the user, (2) when the S2R collection phase starts, (3) right after the user confirms a matched S2R for her S2R description, or (4) when the user has already selected one or more S2Rs suggestions.

\ap implements a shortest-path approach to predict the next S2Rs. First, \ap determines the paths between the current graph state and the corresponding OB state. Then, \ap computes the likelihood score based on the execution model edge weights. 


\ap uses the equation below to compute the score $S_p$ of an $n$-edge path $p = \{w_1, w_2, ... , w_n\}$, with $w_k$ being edge $k$'s weight: 

\hspace{3.2cm}{$S_{p}=\frac{1}{n}{\sum}_{k}{w}_{k}+\frac{1}{n}$}

\noindent  The first term in the sum is the average weight among all path edges and the second term is a factor that favors shorter paths.

Once the paths are ranked by their scores (in descending order), these are modified to include loops, \ie steps that lead to the same app screen (\eg \textit{types} for providing input values). Then, only the first five steps for each path are selected. 
With only the first five steps, all unique paths are kept and only the top-2 paths are saved for being presented to the user. 
The first one is always presented and if the user does not select any of the steps as being the next S2Rs and wants more suggestions, the second path is presented next. Every time the user selects a suggested step as being the next step, the prediction/suggestion process restarts with new predictions.


\subsection{B{\normalsize URT} Implementation}
\label{sec:approach:implemenation_limitations}

\ap is currently implemented as a web application with two major components: the front-end, developed with the React Chatbot Kit~\cite{ReactChatbotKit}, and the back-end, developed with Spring Boot~\cite{SpringBoot}. 
\ap's implementation is tailored for Android applications, however, its underlying techniques are generic enough to be easily implemented for other types of software --- the App Execution Model Data Collection is the only platform-specific part.

To collect the crowdsourced app usage traces for \ap, two computer science students, who did not have knowledge of our studied bugs, were instructed to use the apps' features as they typically would do, and recorded traces that exercise key app features.
Additionally, two of the paper authors recorded sequences simulating app developers who test the apps. These traces were converted and merged into app execution models for each of the studied apps as described in Sec.~\ref{sec:approach:execution_model}. 
In practice, developers can utilize recorded tests, crowdsourced data, or automated app exploration techniques with a ``one-time'' cost for building the app execution model. 



\section{Empirical evaluation design}
\label{sec:evaluation_design}

We conducted two user studies to evaluate: (1) \ap's perceived  usefulness and usability; (2) \ap's intrinsic accuracy in performing bug report element quality verification and prediction; and (3) the quality of the bug reports collected with \ap, compared with reports collected by a template-based bug reporting system. We aim to answer the following research questions (RQs):

\vspace{0.5em}
\begin{enumerate}[label=\textbf{RQ$_\arabic*$:}, ref=\textbf{RQ$_\arabic*$}, wide, labelindent=1pt]\setlength{\itemsep}{0.3em}
	\item \label{rq:usefulness}{\textit{What \ap features do reporters perceive as (not) useful?}}
	\item \label{rq:easeuse}{\textit{What \ap features do reporters perceive as (not) easy to~use?}}
	\item \label{rq:accuracy}\textit{What is the accuracy of \ap in performing bug element quality verification and prediction during the bug reporting process?}
	\item \label{rq:quality}{\textit{What is the quality of the bug reports collected by \ap compared to reports collected by a template-based bug reporting system?}}
\end{enumerate}

To answer the RQs, we selected a set of Android app bugs used in prior research (Sec. \ref{sec:eval_design:dataset}), and asked bug reporters to report these bugs using \ap and to evaluate their experience (Sec. \ref{sec:eval_design:report_collection_and_user_experience}). We analyzed the conversations the reporters had with \ap and measured how accurate \ap was during the reporting process (Sec.~\ref{sec:eval_design:burt_accuracy}). Then, we asked additional participants to report the same bugs with a template-based bug reporting system (Secs.~\ref{sec:eval_design:itracker} and \ref{sec:eval_design:bug_reporting_itrac}), and analyzed the collected bug reports to measure their quality based on bug element correctness (Sec.~\ref{sec:eval_design:quality}). We present and discuss the results in Sec. \ref{sec:results}. Our user studies were approved by an Institutional Review Board (IRB) and conducted remotely due to restrictions related to COVID-19.

\subsection{Apps and Bug Dataset}
\label{sec:eval_design:dataset}

We selected 12 Android app bugs from the bug dataset provided by Cooper \etal~\cite{Cooper:ICSE21}.  
The apps in the dataset support different app domains 
and have been studied in prior research \cite{Bernal:ICSE20,Chaparro:FSE19,Moran2015,Moran2016}. 
The apps are: AntennaPod (APOD) \cite{APOD} -- a podcast manager, Time Tracker (TIME) \cite{TIME} -- a time-tracking app, Android Token (TOK)~\cite{TOK} -- a one-time-password generation app, GnuCash (GNU)~\cite{GNU} -- a personal finances manager, GrowTracker (GROW) \cite{GROW} -- a plant monitoring app, and Droid Weight (DROID) \cite{DROID} -- a personal weight tracking app. 
This dataset provides, for each bug, the APK installer that contains the bug, the description of the incorrect (observed) app behavior (OB), the expected app behavior (EB), and the (minimal) list of the steps to reproduce the bug (S2Rs). 

From the 60 bugs (35 crashes and 25 non-crashes) in Cooper \etal's dataset~\cite{Cooper:ICSE21}, we selected 12 bugs (7 crashes, 1 handled error, and 4 non-crashes) using a stratified random approach (see Table \ref{tab:bug_dataset}). 
We randomly selected two bugs for each of the six apps, ensuring that the bugs represent a variety of bug types that manifest visually on the device (crashes, GUI issues, functional bugs, \etc) and have a diverse number and type of S2Rs (taps, types, swipes, \etc).  
Six bugs contain $5-9$ (minimal) S2Rs, and six bugs contain $10-16$ (minimal) S2Rs (see Table \ref{tab:bug_dataset}).
The 12 bugs are reproducible on a specific web-based Android emulator configuration (virtual Nexus  5X  with  Android  7.0  configured  via the Appetize.io~\cite{Appetize} service). 

\begin{table}[t]
	\setlength\tabcolsep{3.5pt}
	\small
	\centering
	\caption{Apps and bug dataset}
	\label{tab:bug_dataset}
	
	\resizebox{0.92\columnwidth}{!}{%
\begin{tabular}{c|c|c|c}
	\hline
	\textbf{App}           & \textbf{Bug ID} & \textbf{\# of S2Rs} & \textbf{Bug type}                \\ \hline
	\multirow{2}{*}{APOD}  & CC3             & 11                  & Incorrect color in GUI component \\
	& RB              & 5                   & Error message on screen          \\ \hline
	\multirow{2}{*}{DROID} & CC5             & 7                   & Crash                            \\
	& CC6             & 12                  & Crash                            \\ \hline
	\multirow{2}{*}{GNU}   & CC9             & 13                  & Duplicated GUI component         \\
	& RC              & 5                   & Crash                            \\ \hline
	\multirow{2}{*}{GROW}  & CC5             & 10                  & Crash                            \\
	& RC              & 8                   & Crash                            \\ \hline
	\multirow{2}{*}{TIME}  & CC1             & 16                  & GUI component disappears         \\
	& CC4             & 9                   & Crash                            \\ \hline
	\multirow{2}{*}{TOK}   & CC2             & 10                  & Crash                            \\
	& CC7             & 6                   & GUI component does not appear    \\ \hline
\end{tabular}
}
\end{table}

\subsection{RQ$_1$ \& RQ$_2$: B{\normalsize URT}'s User Experience}
\label{sec:eval_design:report_collection_and_user_experience}


We asked participants to report a selected subset of bugs using \ap, and evaluate their experience via an online questionnaire.

\subsubsection{B{\normalsize URT} Bug Reporter Recruitment}
\label{sec:eval_design:reporter_recruitment}

We reached out to 36 potential participants with mixed experience in bug reporting from our personal network, 
who were not involved in or aware of the purpose of this work. 
They were offered a \$15 USD gift card for participation. 
From these, 24 users completed the study and data from six participants was discarded due to low-effort answers, thus resulting in valid responses from 18 participants. 
Four of the six participants did not treat the study seriously, that is, they submitted incomplete reports (e.g., only the OB was reported) and answered all survey questions with the same response. The remaining two participants reported completely different bugs to the ones assigned.
Five participants had not reported a software bug before, nine had reported five or fewer bugs, and the remaining four had reported more than five bugs. 
The participants were unfamiliar with \ap and the selected apps/bugs.

\subsubsection{Bug Assignment and Reporting}
\label{sec:eval_design:bug_reporting}
Each of the 18 participants was randomly assigned to report three bugs (from the 12 selected) with \ap, each bug corresponding to a distinct app. The reporters were instructed to report the bugs in a given (random) order to account for potential learning biases. 
The bug reporting procedure consisted of five tasks which included the users: 
 (i) watching a short instructional video that explained how to use \ap via an example;
 (ii) familiarizing themselves with the apps on the web-based emulator;
 (iii) watching a video demonstrating the observed and expected behavior for each assigned bug  (with annotations to ensure proper understanding); 
 (iv) reproducing the bugs on the web-based emulator; 
 and (v) reporting each bug with \ap. 
 We aimed to control for participant understanding of the bugs so that the effect of potential misunderstandings was minimized.

\subsubsection{B{\normalsize URT}'s User Experience Assessment} 
\label{sec:eval_design:user_experience}

After the participants reported the three assigned bugs, they answered an online questionnaire that was meant to assess \ap's usefulness and ease of use and to obtain feedback for potential improvements to \ap. Table~\ref{tab:survey_questions} shows the questions asked to the participants, which are inspired by the PARADISE~\cite{Hajdinjak2006} evaluation framework.

To address \ref{rq:usefulness}, we focused on evaluating \ap's four main features:
 (1) \ap's app screen suggestions for the OB, EB, and S2Rs;
 (2) \ap's ability to parse and match the OB, EB, and S2R descriptions provided by the user;
 (3) \ap's messages and questions given to the user;
 and (4) \ap's panel of reported S2Rs, which allows the user to visualize and edit the reported S2Rs. 
Questions Q1-Q5 in Table~\ref{tab:survey_questions} aim to address \ref{rq:usefulness} and  used a 5-level Likert scale~\cite{Oppenheim:1992}. We asked the participants to (optionally) provide a justification/rationale for their answers.
Each bug involved multiple screen suggestions, OB/EB/S2R user descriptions, and \ap messages/questions. Questions Q1-Q3 refer to the frequency of these user interactions with \ap.


To address \ref{rq:easeuse}, the reporters assessed \ap's overall ease of use~(Q5) and indicated \ap's specific features that were easy or difficult to use for them~(Q6). Q5 used a used a 5-level Likert scale and Q6 requested an open-ended response. 
The reporters were also asked to indicate additional features they would like to see in \app~(Q7). 
Additional open-ended questions were asked (not shown in Table \ref{tab:survey_questions}) to obtain feedback on how to improve  \ap. 

\begin{table}[t]
	\setlength\tabcolsep{3pt}
	\small
	\centering
	\caption{Questionnaire for evaluating B{\footnotesize URT}'s user experience}
	\label{tab:survey_questions}
	\begin{tabular}{l| m{7.8cm}}
		\hline
		\multicolumn{1}{c|}{\textbf{ID}} & \multicolumn{1}{c}{\textbf{Question}}                                \\ \hline
		Q1                               & How often were \ap's screen suggestions useful?                      \\ 
		Q2 & How often was \ap able to understand your OB/EB/S2Rs?     \\
		Q3                               & How often were you able to understand \ap's messages/questions?  \\ 
		Q4                               & Was \ap's panel of reported steps useful?                            \\ 
		\hline
		Q5                               & How easy to use was \ap overall?                                     \\ 
		Q6 & Which of \ap's features did you find easy/difficult to use?            \\ \hline
		Q7                               & What additional functionality (if any) would you like to see in \ap?
		\\ \hline
	\end{tabular}
\end{table}

\subsection{RQ$_3$: B{\normalsize URT}'s Intrinsic Accuracy}
\label{sec:eval_design:burt_accuracy}

To answer \ref{rq:accuracy}, we analyzed the conversations that the reporters had with \ap to determine: (1) how often \ap was able to correctly match OB/EB/S2R descriptions to the app execution model as confirmed by the reporters;
 and (2) how often the user selected one or more of the suggested app screens as being correct (\ie they match the reporters' OB/EB/S2R descriptions). 
 We computed statistics on the (meta)data that \ap collected from the conversations, such as, the type of messages that \ap asked and the type of user responses (as defined by \ap's Dialogue Manager -- see Sec.~\ref{subsec:dialouge-manager}).

\subsection{RQ$_4$: B{\normalsize URT}'s Bug Report Quality}

We describe the methodology to answer \ref{rq:quality} in this section.

\subsubsection{\itr: A Web Form for Bug Reporting}
\label{sec:eval_design:itracker}

We implemented a web/template-based bug reporting interface, called \itr, using Qualtrics~\cite{qualtrics}. 
\itr offers the same features of professional issue trackers (\eg  GitHub Issues~\cite{github-it} or JIRA~\cite{jira}), for reporting the OB, EB, and S2Rs. 
Specfically, \itr provides text boxes with explicit prompts that ask for the bug summary/title and the OB, EB, and S2Rs. 
In addition, \itr prompts the reporter to provide the S2Rs using a numbered list (via a given template). 
The reporters can write freely  their own bug descriptions in the text boxes and also attach images/files.
We use \itr rather than an existing professional issue tracker to simplify the reporting process for the reporters because they can use \itr without having to log into a service.

\subsubsection{Bug Reporting with \itr}
\label{sec:eval_design:bug_reporting_itrac}

Following the methodology described in Sect. \ref{sec:eval_design:reporter_recruitment}, we recruited 18 more end-users, who did not participate in the \ap study, and asked them to report a subset of bugs using \itr. 
These reporters did not know about \ap, \itr, and the selected apps/bugs, and had a similar bug reporting experience to that of the group who reported the bugs with \ap. 
Five of the new reporters had not previously reported a software bug, eight had reported one to five bugs, and the remaining five had reported more than five bugs.

We assigned the same sets of three bugs used in the \ap study to the new users (trying to match the bug reporting experience) and instructed them to report the bugs using \itr in the same order from before. 
Prior to reporting the bugs, the participants were instructed to: 
(i) familiarize themselves with the apps by using them on the web-based emulator;
(ii) watch a video demonstrating the bugs (with annotations to ensure proper understanding); 
and (iii) reproducing the bugs on the web-based emulator. 

\subsubsection{Measuring Bug Report Quality} 
\label{sec:eval_design:quality}
We estimate the quality of the collected bug reports (via \tool and \itrac) by assessing the correctness of the OB, EB, and S2Rs described in the reports, based on the quality model proposed by Chaparro \etal~\cite{Chaparro:FSE19}.
Three authors manually compared each collected report with the \textit{ground truth scenarios} from Cooper \etal's dataset~\cite{Cooper:ICSE21}, which included correct descriptions of the OB and EB and a minimum viable set of S2Rs. Using this methodology, we computed the following:
 (i) the number of incorrect OB/EB/S2R descriptions; and 
 (ii) the number of missing S2Rs.
To limit the effect of subjective assessments, two authors performed the bug report analysis independently and a third author reviewed the results, reaching consensus among all three in case of discrepancies.
In order to determine how helpful \ap and \itr are for novices or more experienced reporters, we analyzed bug report quality across different levels of bug reporting experience.



\begin{figure}[t]
	\includegraphics[width=\linewidth]{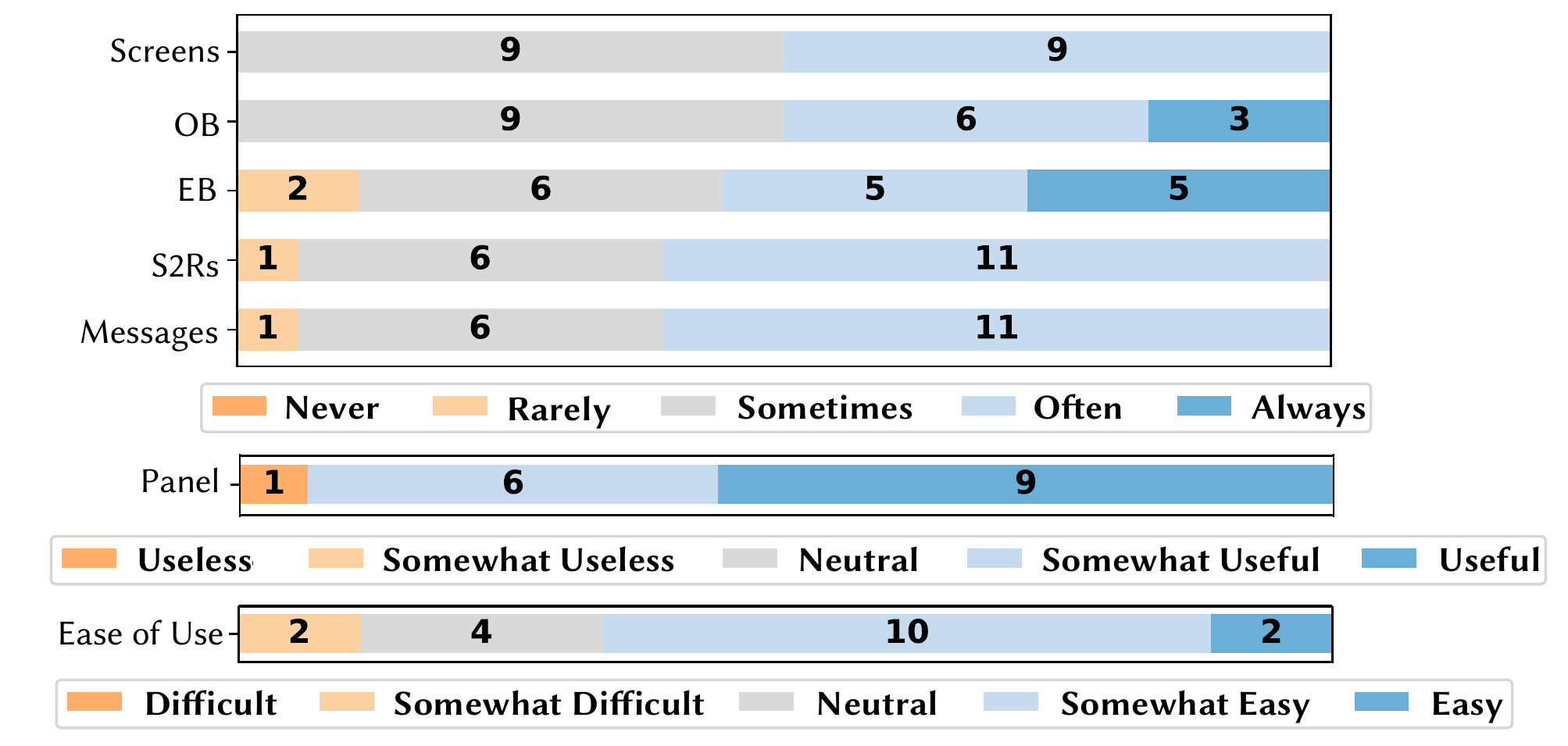}
	\caption{User experience results for  B{\footnotesize URT} (Q1-Q5)}
	\label{fig:usefulness_results}
	\vspace{-0.2cm}
\end{figure}

\section{Results and Analysis}
\label{sec:results}

We present and discuss the results of our evaluation for each RQ. 

\subsection{\ref{rq:usefulness}: B{\normalsize URT}'s Perceived Usefulness}

Fig. \ref{fig:usefulness_results} summarizes the users' answers on: 
(i) their perceived usefulness of \ap's screen suggestions (row labeled \textit{Screens}); 
(ii) \ap's ability to understand the user's OB, EB, and S2R descriptions (rows \textit{OB}, \textit{EB}, and \textit{S2Rs}); 
(iii) how often they were able to understand \ap's messages and questions (row \textit{Messages}); 
(iv) their perceived usefulness of \ap's panel of reported S2Rs (row \textit{Panel});
and (v) \ap's overall ease of use (row \textit{Ease of Use}).

\textbf{App Screen Suggestions}. 
Half of the 18 participants (9) agreed that \ap's app screen suggestions were \textit{often} useful, and the other half (9) agreed they were \textit{sometimes} useful. 
As for their rationales, one participant mentioned that the next S2R screen suggestions \textit{"were useful because they shortened the time it took me to explain how to reproduce the bug"}. 
Other participants highlighted that the suggestions \textit{"were helpful in making sure I was providing the exact steps I wanted to describe"}, or that \ap \textit{``gave very good suggestions when it could figure out which screen had the bug based on the initial report''}. Some of the participants even hoped that \ap can provide suggestions more frequently. These results illustrate the usefulness of \ap's app screen suggestions. 

Some participants noted, though, that \textit{``the suggestions were a little inaccurate''}. 
We found that the inaccuracies stemmed from \ap not being able to recognize/match the user's OB description because of generic wording, without details (\eg \textit{``the app crashed''}). 
Note that the \ap's S2R suggestions are not activated if the OB description is not matched to an app screen, which affected the reporters experience. 
Also, the participants recommended that it would be useful to have suggestions of \textit{``bug triggering screenshots''}, as currently, \ap's screen captures may not show the bug that the user wants to report. 
The participants also found some suggestions confusing because the screen captures for the S2Rs highlight \textit{``non-existent buttons''}. 
This stems from \ap's systematic GUI exploration technique, which can execute events on GUI components such as, layouts or views, which  are often not visible to the user.

\textbf{OB, EB, and S2R Understanding.}
The reporters have a positive overall impression on how often \tool understood their OB, EB, and S2R descriptions. 
Specifically, \ap was able to \textit{often} or \textit{always} (\textit{sometimes}) understand the OB/EB/S2R descriptions of 9/10/11 (9/6/6) participants (out of 18). 
Only two/one participant(s) felt that \ap \textit{rarely} recognized their EB/S2Rs.


Our analysis of the open-ended answers also reveals that some participants were generally satisfied with \ap in terms of bug description understanding. 
This can be seen in comments such as \textit{``I'm quite satisfied with the recognition rate [for the S2Rs], even better than talking to a real agent''}, \textit{``It always understands my description of the OB/EB when I tried to use keywords from apps''}, \textit{``it was kind of easy for burt to understand my (EB) description''}, and \textit{``It can understand me to describe the error behavior''}. 
However, several participants had a less positive perception of \ap's bug description understanding stating that it is \textit{"difficult to match \tool's language"}, they \textit{"need to follow specific pattern"} so that \ap is able to understand, and they \textit{``usually had to paraphrase''} their descriptions. 
These comments stem from our design decision to limit the language that users could use to describe the OB, EB, and S2Rs, and inaccuracies in bug description matching.
 However, we observed, based on the reporters' comments and their conversations with \ap, that the participants learned how to describe the OB/EB/S2Rs using \ap's preferred formats after reporting the first bug. Still, the reporters' main recommendation was to improve \tool's ability to recognize additional 
 vocabulary and ways of phrasing the OB/EB/S2Rs. 

\textbf{B{\small URT}'s Messages and Questions.}
Eleven (of 18) participants \textit{often} understood \ap's messages and questions, while six participants understood them \textit{sometimes}. 
Only in one case, the reporter \textit{rarely} understood the messages/questions. 

The analysis of their  rationales reveals that generally \ap's messages/questions were \textit{``very easy to understand''}. 
One participant wrote that \ap's \textit{``wording was always clear and I could always tell what \ap was asking for''}, also echoed by multiple participants. 
Some participants recommended to improve the messages and questions, as sometimes they were unclear and too similar to each other. 
For example, for \ap's question \textit{``Was this the last S2R that you performed?''}, the participants suggested to clarify which last S2R \ap was referring to.

\textbf{The Panel of Reported S2Rs.}
\ap's panel of reported S2Rs was deemed to be \textit{useful} (\textit{somewhat useful}) by 9 (6) participants. 
Only one participant found that the panel was \textit{somewhat useless}. 
The participants commented that the panel was \textit{``Very useful for visualizing a bug report''}, that \textit{``It was good to see what was getting logged''}, and that it was useful \textit{``as a way for me to review that the reproduction steps I entered are complete``}. 

\textbf{Summary of findings for \ref{rq:usefulness}:} Overall, reporters found \ap's screen suggestions and S2R panel useful. They also had a positive impression of \ap's OB/EB/S2Rs understanding and messages. Improvements are required  for \ap to support additional wording of bug report elements and more accurate suggestions.

\begin{table*}[t]
	\centering
	\caption{Quality assessment results for bug reports (BRs) collected by \ap and \itr}
	\label{tab:results_quality}
	
	\resizebox{0.9\textwidth}{!}{%
	\begin{tabular}{c|cc|cc|cc|cc|cc|cc}
		\hline
		\multirow{3}{*}{\textbf{App-Bug ID}} &
		\multicolumn{2}{c|}{\multirow{2}{*}{\textbf{\# of BRs}}} &
		\multicolumn{2}{c|}{\multirow{2}{*}{\textbf{Avg. \# of S2Rs}}} &
		\multicolumn{2}{c|}{\textbf{Avg. \# (\%) of}} &
		\multicolumn{2}{c|}{\textbf{Avg. \# (\%) of}} &
		\multicolumn{2}{c|}{\textbf{\# of BRs with}} &
		\multicolumn{2}{c}{\textbf{\# of BRs with}} \\
		&
		\multicolumn{2}{c|}{} &
		\multicolumn{2}{c|}{} &
		\multicolumn{2}{c|}{\textbf{incorrect S2Rs}} &
		\multicolumn{2}{c|}{\textbf{missing S2Rs}} &
		\multicolumn{2}{c|}{\textbf{incorrect OB}} &
		\multicolumn{2}{c}{\textbf{incorrect EB}} \\ \cline{2-13} 
		&
		\textbf{\itr} &
		\textbf{\ap} &
		\textbf{\itr} &
		\textbf{\ap} &
		\textbf{\itr} &
		\textbf{\ap} &
		\textbf{\itr} &
		\textbf{\ap} &
		\textbf{\itr} &
		\textbf{\ap} &
		\textbf{\itr} &
		\textbf{\ap} \\ \hline
		APOD-CC3  & 5 & 5 & 4.6 & 7.4  & 0.6 (16.7\%) & 0.6 (9.7\%)  & 6.4 (58.2\%) & 3.4 (30.9\%) & 0 & 1 & 0 & 3 \\
		APOD-RB   & 4 & 4 & 3.3 & 4.8  & 0.0 (0.0\%)  & 1.8 (30.6\%) & 1.0 (20.0\%) & 1.0 (20.0\%) & 1 & 1 & 1 & 0 \\ \hline
		DROID-CC5 & 6 & 6 & 4   & 4.3  & 0.3 (11.1\%) & 0.5 (10.0\%) & 1.3 (19.0\%) & 0.7 (9.5\%)  & 1 & 2 & 1 & 1 \\
		DROID-CC6 & 6 & 6 & 4.5 & 8.5  & 1.3 (44.4\%) & 1.2 (13.7\%) & 5.0 (41.7\%) & 2.0 (16.7\%) & 0 & 3 & 1 & 1 \\ \hline
		GNU-CC9   & 5 & 5 & 6.4 & 10.2 & 0.8 (26.7\%) & 0.2 (2.5\%)  & 4.8 (36.9\%) & 3.2 (24.6\%) & 1 & 0 & 1 & 0 \\
		GNU-RC    & 3 & 3 & 4.7 & 4.3  & 0.0 (0.0\%)  & 0.3 (6.7\%)  & 0.0 (0.0\%)  & 0.0 (0.0\%)  & 0 & 1 & 0 & 0 \\ \hline
		GROW-CC5  & 4 & 4 & 4.8 & 7.5  & 1.3 (28.1\%) & 0.0 (0.0\%)  & 4.5 (45.0\%) & 3.5 (35.0\%) & 0 & 0 & 0 & 0 \\
		GROW-RC   & 4 & 4 & 5.8 & 7.5  & 1.0 (30.0\%) & 0.5 (7.1\%)  & 1.8 (21.9\%) & 1.5 (18.8\%) & 1 & 3 & 1 & 0 \\ \hline
		TIME-CC1  & 5 & 5 & 7.8 & 10.4 & 1.0 (24.0\%) & 0.2 (2.9\%)  & 6.6 (41.3\%) & 6.2 (38.8\%) & 0 & 1 & 0 & 0 \\
		TIME-CC4  & 4 & 4 & 4.3 & 8    & 1.0 (24.4\%) & 0.3 (5.0\%)  & 3.0 (33.3\%) & 1.3 (13.9\%) & 1 & 2 & 2 & 1 \\ \hline
		TOK-CC2   & 4 & 4 & 4.8 & 10.3 & 0.3 (8.3\%)  & 0.8 (6.8\%)  & 2.5 (25.0\%) & 0.5 (5.0\%)  & 2 & 2 & 0 & 0 \\
		TOK-CC7   & 4 & 4 & 5   & 5.8  & 0.5 (16.7\%) & 0.3 (3.6\%)  & 1.5 (25.0\%) & 0.8 (12.5\%) & 1 & 0 & 1 & 0 \\ \hline
		\textbf{Overall} &
		\textbf{54} &
		\textbf{54} &
		\textbf{5} &
		\textbf{7.5} &
		\textbf{0.7 (20.4\%)} &
		\textbf{0.6 (8.3\%)} &
		\textbf{3.4 (32.0\%)} &
		\textbf{2.1 (19.4\%)} &
		\textbf{8} &
		\textbf{16} &
		\textbf{8} &
		\textbf{6} \\ \hline
	\end{tabular}
}

\end{table*}

\subsection{\ref{rq:easeuse}: B{\normalsize URT}'s Perceived Ease of Use}

Twelve reporters indicated \ap was either \textit{easy} or \textit{somewhat easy} to use. Four reporters  were neutral, while two reporters expressed it was \textit{somewhat difficult} to use (see \textit{Ease of use} in Fig. \ref{fig:usefulness_results}). 

We analyzed the reporter responses regarding which of \ap's features they found easy/difficult to use. 
In general, the participants expressed that \ap's GUI \textit{``is really helpful''}, 
\textit{``concise''}, and \textit{``easy to use and understand''}. Multiple reporters indicated that selecting \ap's app screen suggestions was easy to use and some of them were very enthusiastic about them. One reporter mentioned that \textit{"I liked the screenshots a lot, very easy to report the process to reproduce a bug"}. Other reporters expressed that \textit{"The suggestions \& confirmations were very easy to use. When it had the right idea, confirming it was just a matter of clicking a button"}, and that \ap \textit{``guides the user to provide a "step-by-step" view''}.  The panel of reported steps was easy \textit{``to explore''} and it was easy to \textit{``remove events''} from it. 

The main reason behind usage difficulties was the limited vocabulary that \ap understands, also observed before for \ref{rq:usefulness}. 
The reporters recommended to let the users upload their own screen captures when \ap is unable to attach screens to the user's bug descriptions, and the ability to delete/modify \textit{any} step.

Finally, for both \ref{rq:usefulness} \& \ref{rq:easeuse}, we found no notable differences in \ap's perceived usefulness and ease of use between different levels of user's bug reporting experience.

\subsection{\ref{rq:accuracy}: B{\normalsize URT}'s Intrinsic Accuracy}

We analyzed the 54 conversations that reporters had with \ap to determine how often \ap was able to correctly (1) match OB/EB/S2R descriptions to the execution model, and (2) suggest relevant OB/S2R app screens to the reporters.

\textbf{OB Reporting.} 
We found that in 3 of 54 conversations (5.5\%), \ap was able to match the reporter's OB description  to the correct screen that showed or triggered the bug, as confirmed by the reporter during the conversation. In 35 of 54 conversations (64.8\%), \ap matched the OB description to multiple app screens. In those cases, \ap suggested the top-5 matched screens so that the reporter selected the one s/he was referring to. In 29 of these 35 reports (80\%), the reporter selected one of the suggested screens, while in the remaining 6, the suggested screens were irrelevant. For the remaining 16 of the 54 conversations (29.6\%), \ap was not able to match the OB description with any app screen because of incorrect OB wording from the user and inaccuracies in \ap's message parser and processing. 
Overall, \ap was able to correctly match their OB descriptions in 32 of 54 of the conversations (59.3\%).

\textbf{EB Reporting.}
As described in Sect.~\ref{sec:approach:matching}, \ap can only match the reporter's EB description when there is a matched/selected OB screen. Otherwise, \ap collects the EB description from the user as is. 
In the 32 cases when \ap can verify EB quality, \ap was able to match the EB against the OB screen in 17 cases (53.1\%) without having to ask the reporter for confirmation. 
In 6 of the 32 cases (18.8\%), the users confirmed the matched OB screen when \ap asked them about that.
In the remaining 9 cases (28.1\%), \ap was not able to parse the provided EB description.
 
\textbf{S2R Reporting.}
\ap matched a written S2R with a step from the execution model 205 times in total across the 54 conversations (3.8 times per conversation on avg.). In 157 of these cases (76.6\%), \ap was able to match S2Rs correctly. 
\ap predicted and suggested the next S2Rs in 146 cases (4.6 times per conversation on avg.) for the 32 conversations where there was a matched/selected OB screen. 
We found that the reporters selected 1.6 of the 3.9 suggested S2Rs (on avg.) in 91 cases (62.3\%). 
In 13 of the 32 conversations, the reporter always selected S2Rs from the suggested list, meaning at least one suggestion was correct.
In all the 54 conversations, \ap asked the user to rephrase their S2Rs 176 times (3.9 times per conversation on avg.). 
We found that in at least 59 of these cases (33.5\%), the user made a mistake or described the step incorrectly (\eg \textit{``incorrect result''} or \textit{``no more steps''}). 

\textbf{Summary of findings for \ref{rq:accuracy}:}
The results support the users' ratings (\ref{rq:usefulness}) on how often \ap's OB/S2R screen suggestion were useful and how often \ap was able to understand the user's OB/EB/S2R descriptions.
The accuracy assessment revealed cases where \ap's struggles to parse and match the users' descriptions, however,
\ap is able to continue with rephrasing prompts.
The overall accuracy indicates that the techniques we used in building \ap's components are adequate.
Improvements are planned for future work to improve \ap's accuracy.

\subsection{\ref{rq:quality}: Bug Report Quality}


Table \ref{tab:results_quality} summarizes the quality measures of the  $54\times2=108$ bug reports, collected with \itr and \ap, for the 12 bugs in our dataset (each bug is reported in 3 to 6 reports). 

\textbf{S2R Quality.} Overall, as shown in Table \ref{tab:results_quality}, \ap reports contain fewer incorrect S2Rs than \itr reports on avg. (8.3\% vs. 20.4\%) and fewer missing S2Rs (19.4\% vs. 32\%), compared to the ground-truth scenarios of the 12 bugs. 
We performed an analysis to verify whether there there statistically significant differences between \ap and \itrac on the percentage of incorrect and missing S2Rs. 
	We applied the Wilcoxon signed-rank test~\cite{Hollander2013} and Cliff's delta (CD)~\cite{Cliff2014} on the results, across the 12 bugs (at 95\% confidence level), since we have paired ordinal measurements (for each bug) that do not necessarily follow normal distributions.
We found that \ap's bug reports have fewer incorrect ($p=0.0261$) and fewer missing steps ($p=0.0025$) than \itrac's reports, with a large effect size (CD $=0.5$ and $0.527$, respectively).

The main reasons for incorrect S2Rs are generic/unclear step wording (4 in \ap and 36 \itr reports), duplicate S2Rs (13 in \ap reports, zero in \itr reports), and extra S2Rs (10 in \ap and one in \itr reports). 
Examples of steps with unclear/generic wording include \textit{``Add comment``} or \textit{``I searched for tech``}, where the user either refers to high-level app features, which map to multiple steps that are not explicit, or does not specify which GUI components should be used and/or which action should be applied on them.  Extra S2Rs are irrelevant reported steps (\eg ``\textit{I did nothing else}``). 
We identified two main reasons for duplicate S2Rs: (1) user mistakes; and (2) duplicate app screens suggested by \ap and selected by the users. 
The latter stems from the design of \ap's execution model that considers structural variations of the same screen as different screens (see Sec.~\ref{sec:approach:execution_model}). 
An example is when the users employ different keyboard layouts (\eg numeric vs. alphanumeric) to enter input values on the same screen.

\textbf{OB/EB Quality.} More \ap reports have an incorrect OB description compared to \itr reports (16 vs. 8 out of 54 reports), while a comparable number of \ap and \itr reports have an incorrect EB description (8 vs. 6). 
We found that there is no statistically significant difference between the number of \ap and \itrac bug reports with incorrect expected behavior ($p=0.1586$), with a small effect size (CD $=0.222$) in favor of \ap. Fewer \itrac reports than \ap reports have an incorrect observed behavior ($p=0.0352$), with a medium effect size (CD $=0.361$).

The incorrect OB/EB descriptions (in 18 \ap reports and 10 \itr reports total) occurred either because the participants did not provide enough details about the bug (\eg \textit{``the app crashed''}) or they described their inability to perform an action rather than describing the bug itself (\eg \textit{\textit{``I can't add/delete a comment''} vs. \textit{``Crash when trying to  add/delete a comment''}}). 

For the 18 \ap reports, we found that, in 14 cases the users described the OB/EB incorrectly to begin with and \ap correctly prompted them to rephrase them. 
Nonetheless, they still reported an incorrect OB/EB. 
In four cases, \ap accepted the incorrect OB/EB, and in only three of the cases, \ap prompted incorrect OB/EB reporting after the user correctly described them. 
This is mainly due to \ap's current limitation on the OB/EB wording. 

\textbf{Summary of findings for \ref{rq:quality}:} Overall, \ap bug reports contain higher-quality S2Rs than \itr bug reports, and comparable EB descriptions. The results indicate that improvements to \ap are needed to better collect OB descriptions from the reporters.

\begin{table}[]
	\centering
	\setlength\tabcolsep{2pt}
	\caption{S2R quality by bug reporting experience}
	\label{tab:results_quality_exp}
	\resizebox{\columnwidth}{!}{%
	\begin{tabular}{c|cc|cc|cc|cc}
		\hline
		\multirow{2}{*}{\textbf{Reporting}} &
		\multicolumn{2}{c|}{\multirow{2}{*}{\textbf{\# of BRs}}} &
		\multicolumn{2}{c|}{\textbf{Avg. \# of}} &
		\multicolumn{2}{c|}{\textbf{Avg. \% of}} &
		\multicolumn{2}{c}{\textbf{Avg. \% of}} \\
		&
		\multicolumn{2}{c|}{} &
		\multicolumn{2}{c|}{\textbf{S2Rs}} &
		\multicolumn{2}{c|}{\textbf{incorrect S2Rs}} &
		\multicolumn{2}{c}{\textbf{missing S2Rs}} \\ \cline{2-9} 
		\textbf{experience} &
		\textbf{\itr} &
		\textbf{\ap} &
		\textbf{\itr} &
		\textbf{\ap} &
		\textbf{\itr} &
		\textbf{\ap} &
		\textbf{\itr} &
		\textbf{\ap} \\ \hline
		Novice &
		15 &
		15 &
		3.5 &
		6.7 &
		33.6\% &
		6.7\% &
		45.6\% &
		31.3\% \\
		Intermediate &
		24 &
		27 &
		5.2 &
		7.5 &
		20.1\% &
		11.5\% &
		35.5\% &
		20.0\% \\
		Experienced &
		15 &
		12 &
		6.1 &
		8.6 &
		7.6\% &
		3.2\% &
		12.8\% &
		3.2\% \\ \hline
		\textbf{Overall} &
		\textbf{54} &
		\textbf{54} &
		\textbf{5} &
		\textbf{7.5} &
		\textbf{20.4\%} &
		\textbf{8.3\%} &
		\textbf{32.0\%} &
		\textbf{19.4\%} \\ \hline
	\end{tabular}

}
\end{table}

\textbf{Novice vs. Experienced Bug Reporters.}
Our original expectation was that \ap would help novice reporters more than \itr, as the experienced reporters likely used template-based reporting systems before.

We compared the quality of the bug reports across different levels of user's bug reporting experience. 
While we did not observe notable differences in terms of OB/EB quality, we found differences in S2R quality, which we discuss.
Table~\ref{tab:results_quality_exp} shows the S2R quality results for three groups: novice bug reporters (with \textit{no} prior reporting), intermediate reporters (who had reported 1-5 bugs), and experienced reporters (who had reported 6+ bugs).

Regarding incorrect S2Rs, experienced and intermediate reporters produced about twice as many incorrect S2Rs with \itr, compared to \ap (33.6\% vs. 6.7\%, and 20.1\% vs. 11.5\% on avg., respectively).
At the same time, novices produced about five times more incorrect steps with \itr than with \ap (7.6\% vs. 3.2\% on avg.). 
This indicates that \ap helps novices most to avoid incorrect S2Rs.

Table~\ref{tab:results_quality_exp} tells a different story for missing S2Rs. 
Novices and intermediate reporters missed $\approx$1.5 times fewer S2Rs with \ap, compared to \itr, while experienced reporters missed four times fewer S2Rs with \ap. 
Surprisingly, this indicates that \ap helps experienced reporters most to avoid missing steps. 

We do not speculate on the reasons behind these observations, as more in-depth studies are needed for proper explanations.

\section{Limitations and Threats to Validity}
\label{sec:limitations_threats}

Before \ap is deployed for use, either systematic app exploration data or crowdsourced app usage data needs to be collected to construct the app execution model. The evaluation results indicate that \ap performs reasonably well with the data collected by \crashscope and only four people. However, we expect that additional data (more covered states and scenarios) would improve \ap's quality verification of reported elements and screen/step suggestions, enabling the reporting of different bug types, under a variety of reproduction scenarios. To confirm our expectations, additional studies are needed for future work.

\ap is evaluated in a lab setting where reporters were exposed to the bugs through videos, rather than letting them find the bugs while using the apps, as users would do in real life. As in prior studies~\cite{Chaparro:FSE19,Moran2015}, we adopted this setting mainly to reduce participant effort and fatigue. To address the lack of knowledge about the apps/bugs, we instructed the users to get familiar with the apps by using them and with the bugs by reproducing them on the emulator before they reported the bugs. We addressed potential bug misunderstandings via 2/3-word annotations added to the videos.

A diverse group of reporters participated in the studies, who have different levels of bug reporting experience. Since we offered the reporters a monetary incentive for their participation and some of them are students from our institution(s), they may have been motivated to diligently provide high-quality bug reports, which may not necessarily be the case in a real-life scenario. However, we expect this factor to have a minimal impact on the results since (1) we used the same procedure to recruit both \ap and \itr users, and (2) the bug reporting experience in both reporter groups are almost the same (only two \itr users have a different experience). 

Our evaluation did not consider how easy or difficult it is (for developers) to understand and reproduce the \itr and \ap bug reports. Instead, we focused on assessing bug report quality, as  done by prior work~\cite{Chaparro:FSE19}. Assessing bug report understanding and reproduction is in our plans for future work. Additionally, we did not account for the complexity of the bugs in our dataset. However, we selected bugs of diverse types and distributions of the S2Rs.
Our future work will investigate how bug complexity affects \ap. 

Finally, given the relatively expensive nature of our evaluation, we limited it to 12 bugs from 
six apps, reported by 36 participants, which affects the external validity of our conclusions. A larger evaluation, possibly performed on a larger sample of apps, bugs, and participants is in our plans for future work.

\section{Related work}
\label{sec:related_work}

We discuss \ap's advancements in relation to prior work.

\textbf{Issue/Bug Reporting Systems}. A variety of systems currently enable end-users and developers to manually report software bugs, namely, issue/bug trackers (\eg GitHub Issues~\cite{github-it} or JIRA~\cite{jira}), built-in bug reporting interfaces in desktop and web apps (\eg Google Chrome~\cite{ChromeReportIssue}), in-app bug reporting frameworks (\eg BugSee \cite{bugsee}), app stores ~\cite{GooglePlayAppReviews,AppStoreAppReviews}, and Q\&A platforms~\cite{bhatia2022study}. These systems typically consist of web/GUI forms (with text-based templates) that allow reporters to provide bug descriptions, indicate bug/system metadata, and attach relevant files. Some of these systems collect technical information (\eg configuration parameters) and offer screen recording that enable graphical bug reporting. 

While existing systems provide features that facilitate bug reporting, they offer limited guidance to bug reporters, lack quality verification of bug report information, and do not provide concrete feedback on whether this information is correct and complete. These are some of the main reasons for having low-quality bug reports, which have important repercussions for developers~\cite{Zimmermann2009,Zimmermann2010}.

Researchers have explored improving bug reporting interfaces,
as we do in this work. Moran \etal~\cite{Moran2015} proposed \fusion, a web-based system that allows the user to report the S2Rs graphically by selecting (via dropdown lists) images of the GUI components 
and actions (taps, swipes, \etc) that can be applied on them. More recently, Fazzini~\etal proposed \ebug~\cite{Fazzini:TSE22}, a mobile app bug reporting system similar to \fusion that suggests potential future S2Rs to the reporter while they are writing them.
Record-and-replay tools~\cite{Moran:MobileSoft17,Gomez2013,Hu:OOPSLA15,Qin2016} 
offer the ability to record user actions during app usage (\eg when a bug is found) and replay them later.

\ap offers two main advancements over prior techniques like \fusion. 
First, \ap was designed to support end-users with little or no bug reporting experience. 
For example, \fusion was not created to specifically cater to end-users, as inexperienced users found it \textit{more difficult} to use as compared to  alternatives~\cite{Moran2015}. 
Second, whereas past systems helped to provide structured mechanisms to facilitate the reporting process (\eg through drop-down selectors) they do not offer \textit{interactive} assistance when reporting a bug. 
\ap offers such interactivity through its automated suggestions, real-time quality assessment, and prompts for information clarification. 

\textbf{Bug Report Quality Analysis.} 
Surveys and interviews with developers and end-users~\cite{Zimmermann2010,Laukkanen2011,Sasso2016} have identified the observed software behavior (OB), the expected behavior (EB), and the steps to reproduce (S2Rs) the bugs as essential bug report elements for developers during bug triage and resolution. Unfortunately, such elements are often missing, unclear, or ambiguous, as indicated by numerous studies and developers~\cite{ErfaniJoorabchi2014,Zimmermann2012,Breu2010,Guo2010,Zimmermann2010,Fazzini2018,Karagoez2017,GitHub2016}, which have a negative impact on bug report management tasks.

In consequence, researchers have proposed techniques to better capture and manage high-quality information in bug reports.
Prior work~\cite{Bettenburg2008a,Davies2014,Chaparro:FSE17,Song:FSE20,Zhao2019,zhao2022recdroid+,Zimmermann2010} proposed ways to automatically identify different essential elements in bug reports (\eg S2Rs \cite{zhao2019automatically,liu2020automated}), analyze their quality, and give feedback to reporters about potential issues in them. In particular, Zimmermann \etal~\cite{Zimmermann2010} proposed an approach to predict the quality level of a bug report based on factors such as readability or presence of keywords. 
Hooimeijer \etal~\cite{Hooimeijer2007} measured quality properties of bug reports (\eg readability) to predict when a report would be triaged. Zanetti \etal~\cite{Zanetti2013} proposed an approach based on collaborative information to identify invalid, duplicate, or incomplete bug reports. Imran \etal~\cite{Imran2021} proposed an approach to suggest follow-up questions for incomplete reports. Song \etal~\cite{Song:FSE20,Chaparro:FSE17} proposed a technique to detect when the OB, EB, and S2Rs are absent in submitted bug reports.  Chaparro \etal~\cite{Chaparro:FSE19} evaluated the quality of the S2Rs in bug reports through the \euler tool, which integrates dynamic app analysis, natural language processing, and graph-based approaches. 

Our work builds upon prior research for the automated quality verification of bug descriptions by developing quality checks for new types of bug elements (\ie OB/EB) and by designing dialogue flows capable of guiding the user during the bug reporting process.

\section{Conclusions}

\ap is a task-oriented chatbot for interactive Android app bug reporting. 
Unlike existing bug reporting systems, \ap can guide end-users in reporting essential bug report elements (\ie OB, EB, and S2Rs), provide instant feedback about problems with this information, 
and produce graphical suggestions of the elements that are likely to be reported. 

Eighteen end-users reported 12 bugs from six Android apps and reported that, overall, \ap's guidance and automated suggestions/clarifications are accurate, useful, and easy to use.
The resulting bug reports are higher-quality than reports created via \itr, a template-based bug reporting system, by other 18 reporters. 
Specifically, \ap reports contain fewer incorrect and missing reproduction steps compared to \itr reports.
We observed that \ap is most helpful to novice reporters for avoiding incorrect S2Rs.
Surprisingly, \ap seems to be most useful to experienced reporters for avoiding missing reproduction steps. 

The reporters provided feedback for refining the supported dialog, by including support additional wordings to describe the OB, EB, and S2Rs.
The studies also revealed areas of improvement for \ap with respect to the verification of the reported elements.



\section*{acknowledgements}

We thank all the study participants for their time and feedback. This work is supported by the NSF grants: CCF-1955837 and CCF-1955853. Any opinions, findings, and conclusions expressed herein are the authors and do not necessarily reflect those of the sponsors.


\balance
\bibliographystyle{ACM-Reference-Format}
\bibliography{references}

\end{document}